\documentclass[prl,amsmath,amssymb,twocolumn,superscriptaddress,showpacs]{revtex4-1}
\usepackage{bm}
\usepackage{amssymb}
\usepackage{colordvi}
\usepackage{graphicx}
\usepackage{color}
\usepackage{hyperref}
\usepackage{ulem}
\usepackage{dcolumn}

\newcommand{\beq}{\begin{equation}}
	\newcommand{\eeq}{\end{equation}}
\newcommand{\bea}{\begin{eqnarray}}
	\newcommand{\eea}{\end{eqnarray}}

\begin{document}
	
	
	\title{Topological light guiding and trapping via shifted photonic crystal interfaces}
	
	\author{Zi-Mei Zhan}
	\affiliation{School of Physical Science and Technology, Guangxi Normal University, Guilin 541004, China}
	
	\author{Peng-Yu Guo}
	\affiliation{School of Physical Science and Technology, Guangxi Normal University, Guilin 541004, China}
	
	\author{Wei Li}
	\affiliation{School of Physical Science and Technology, Guangxi Normal University, Guilin 541004, China}
	
	\author{Hai-Xiao Wang}%
	\email{hxwang@gxnu.edu.cn}
	\affiliation{School of Physical Science and Technology, Guangxi Normal University, Guilin 541004, China}
	
	\author{Jian-Hua Jiang}%
	\email{jianhuajiang@suda.edu.cn}
	\affiliation{School of Physical Science and Technology, \& Collaborative Innovation Center of Suzhou Nano Science and Technology, Soochow University, Suzhou 215006, China}
	\affiliation{Suzhou Institute for Advanced Research, University of Science and Technology of China, Suzhou 215123, China}
	
	\date{\today}
	
	\begin{abstract}
		The exploration of topological states in photonic crystals have inspired a number of intriguing discoveries, which in turn provide new mechanisms for the manipulation of light in unprecedented ways. Here we show that light can be effectively guided and trapped at the shifted photonic crystal interfaces (SPCIs). The projected band gap of SPCIs, which depends on the shift parameter, is characterized by a Dirac mass. Interestingly, the SPCI with zero Dirac mass is a glide-symmetric waveguide featured with gapless interface states that exhibit excellent transmission performance even in the presence of disorders and sharp corners. Moreover, placing two SPCIs with opposite Dirac mass together results in a photonic bound state due to the Jackiw-Rebbi theory. Our work provides an alternative way towards the design of ultracompact photonic devices such as robust waveguides and cavities as well as the cavity-waveguide coupled systems that can serve as high-performance building blocks of miniature integrated topological photonic circuits. 
	\end{abstract}
	
	\maketitle
	
	
	{\it Introduction.---} The past decades have witnessed the rapid development of various topological states in photonic systems~\cite{topo1,topo2,topo3}. Typical photonic topological phases, including photonic quantum anomalous Hall effects~\cite{QAH1,QAH2,QAH3,QAH4}, photonic Floquet topological insulators~\cite{floquet1,floquet2,floquet3}, and photonic quantum spin Hall insulators~\cite{QSH1,QSH2,QSH3,QSH4,QSH5,QSH6}, support topologically protected edge states that are appealing in guiding light with suppressed back scattering. Recently, the concept of Wannier-type higher-order topological insulators~\cite{WHOTI1} was proposed to give rise to topological corner or hinge states, which provide an alternative way to trap light in lower-dimensional boundaries~\cite{WHOTI2,WHOTI3,WHOTI4,WHOTI5,WHOTI6,WHOTI7,WHOTI8,WHOTI9,WHOTI10}. The basic idea is to deal with the Wannier configuration in the unit cells through the lattice deformation. For example, the breathing kagome lattice generates nontrivial Wannier configuration, and manifests its higher-order topology in the gapped edge states and in-gap corner states~\cite{WHOTI4,WHOTI9,WHOTI10}. In fact, placing photonic systems with different Wannier configurations provides an effective way to guide and trap light~\cite{Wannier1,Wannier2,Wannier3,Wannier4,Wannier5}, which may found potential applicationss on photonic devices like topological lasers~\cite{WHOTI5} and rainbow light trappings~\cite{Wannier6}. On the other hand, it is recognized that glide symmetry, a composite symmetry operator that consists of mirror and translation operation, is beneficial for engineering the topological band structures. For example, the glide symmetry provides a powerful tool to synthetic Kramers degeneracy in topological classical wave systems~\cite{glidetopo1,glidetopo2,glidetopo3,glidetopo4,glidetopo5}. In particular, it was shown that glide-symmetric interface can induce topological Wannier cycles from which gapless spin-Hall-like interface states emerge in the bulk band gap~\cite{glidetopo2}. Very recently, a glide-symmetric phononic crystal interface that supports wide-bandwidth, single-mode topological acoustic waves is proposed~\cite{gswg1,gswg2}, and is suggested to have potential applications on the extremely sensing and isolation~\cite{gswg3}. It is believed that such a topological acoustic wave is originated from the Wannier configuration on the interface rather than the bulk topology, which is differ from conventional topological insulator~\cite{Wannier6}. Nevertheless, the glide-symmetric photonic crystal (PC) interface remains unexplored and it is unknown whether or not such an interface can support any localized mode similar to that in the Wannier-type higher-order topological corner states. 
	
	In this letter, we demonstrate that shifted photonic crystals interfaces (SPCIs) can support both gapless interface states and in-gap bound states, which can have potential applications on robust light guiding and trapping. When the shift parameter equals to half lattice constant, the SPCI with a nontrivial Wannier configuration support gapless interface states that can be utilized as a waveguide featured with excellent transmission performance even in the presence of sharp corner. Moreover, when two SPCIs with opposite shift parameters are placed together, the in-gap localized states emerged as the Jackiw-Rebbi solitons of the gapped interface states. Finally, we discuss the coupled waveguide-cavity systems formed by the SPCIs-induced waveguide and localized states, showing that such systems can serve as elementary units for tunable, high-performance integrated photonic circuits.
	
\begin{figure}[htbp]
		\centering\includegraphics[width=\columnwidth]{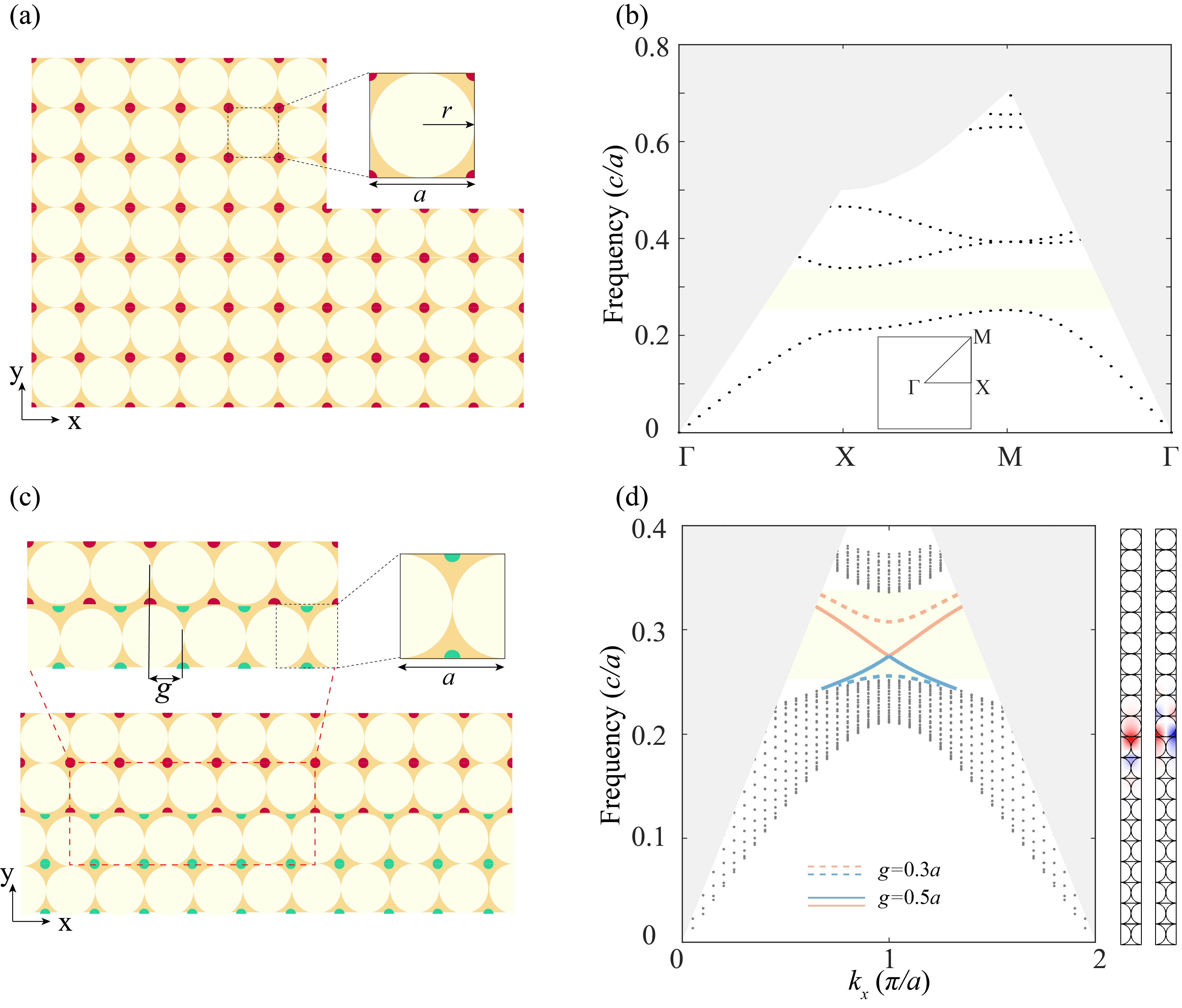}
		\caption{(a) Schematic of two-dimensional PCs consisting of circular air holes with $r=0.5a$ arrayed in square lattice. The positions of the Wannier centers for the first band are indicated by the red dots. (b) The photonic band structure for the proposed PCs. Inset: the first Brillouin zone. (c) Schematic of a SPCI that formed by moving the lower half PC with shift parameter $g$. The red and green dots refer to the position of Wannier center of the PCs upper and below the SPCIs with $g=a/2$. (d) Left panel: the projected band structure of SPCIs, where the solid (dashed) lines refer to interfaces states with $g=\pm 0.5a (g=\pm 0.3a)$. Right panel: the eigen electric fields of interface states at $k_x=\pi/a$ with odd and even parities.}
		\label{Fig_1}
	\end{figure}

	{\it SPCIs.---}We start from a two-dimensional PCs with circular air holes arrayed in square lattice in a dielectric background [see Fig.~\ref{Fig_1}(a)]. The radius of each circular air hole is $r=0.5a$, where $a$ is lattice constant. Throughout this work, we use silicon ($\epsilon=11.9$) as the dielectric background, and consider only the transverse-magnetic (TM) mode. This setup is simple and effective, as shown in the results below. Moreover, the same design can be applied to a broad frequency range, from optical frequencies to microwave frequencies. All simulations here are carried out with the commercial software COMSOL Multiphysics. We first present the photonic band structure in Fig.~\ref{Fig_1}(b), where the grey and light yellow regions refer to the states above light cone and complete band gap, respectively. Since we only focus on the first band gap, it is worthy to point out that the Wannier center for the first band are located at the corners of the primitive cell [red dots in Fig.~\ref{Fig_1}(a)]. Generally, the Wannier center position is connected to the bulk topological polarization and hence can be employed as a bulk topological index for crystalline insulators [see calculation details of the bulk topological polarization in Supplementary Material].  
	
\begin{figure}[htbp]
		\centering\includegraphics[width=\columnwidth]{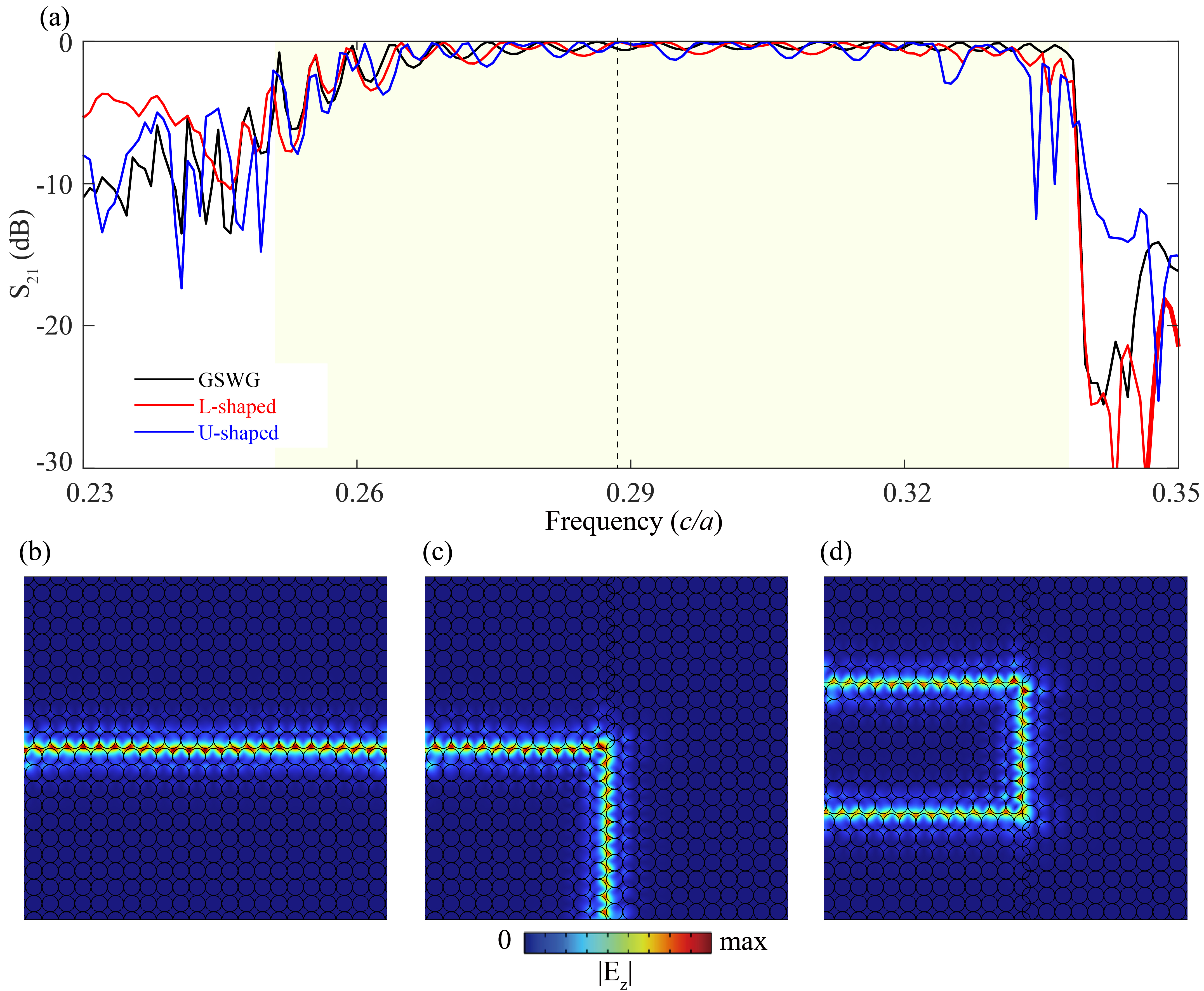}
		\caption{(a) Transmission of GSWG (black line), L-shaped (red line) and U-shaped (blue line) SPCI versus frequency. light yellow refers to the bulk band gap. (b-d) The electric field patterns of GSWG, (c) L-shaped and (d) U-shaped SPCIs, respectively, at a frequency of $0.289(c/a)$. }
		\label{Fig_2}
	\end{figure}

	To illustrate the SPCI, a geometry parameter $g$ is employed to describe the relative displacement of the PCs between upper and lower interfaces. As indicated by dashed rectangular box in Fig.~\ref{Fig_1}(c), a SPCI forms by moving the lower half PC along $x$-directions with a distance $g$ while the upper half PC keeps unchanged. In general, the parameter $g$ ranges from $-a/2$ to $+a/2$ due to the periodicity of the PC. Here, the positive and negative signs refer to the rightward and leftward moving directions, respectively. Obviously, the initial PCs represent the situation with $g=0$. For $g=\pm a/2$, the SPCIs can be regarded as an interface formed by square PCs with two different Wannier configurations. As depicted by the green and red dots in Fig.~\ref{Fig_1}(c), the Wannier centers in the lower half PC is shift compared to the upper half PC. Remarkably, such a simple shift results in two interface bands crossing linearly at $k_x=\pi/a$ [see the solid colored lines in Fig.~\ref{Fig_1}(d)]. To understand the formation of such a band crossing (also known as a Dirac point), we construct an anti-unitary operator $\Theta_x= G_x\star T$, where $G_x: (x,y) \rightarrow (x \pm a/2,-y)$ and $T$ are glide operator and time-reversal symmetry, respectively. Hence, acting twice $\Theta_x$ on a photonic Bloch states give additional phase factor $e^{-ik_x\cdot a}$. In other words, $\Theta_x$ transforms $(k_x,k_y)$ into $(-k_x,k_y)$ and we have 
	\begin{equation}
		{\Theta_x}^2=e^{-ik_xa}|_{k_x=\pi/a}=-1,
		\label{Kramers}
	\end{equation} 
	which, as an analog of the Kramers theorem fermions, guarantees that all bands are doubly degenerate and forming pairs at $k_x=\pi/a$. Such gapless interface states protected by the synthetic Kramers theorem exhibit  excellent performance in guiding waves [also see in Fig.~\ref{Fig_2}] and have been utilized as the glide-symmetric waveguide (GSWG)~\cite{gswg1,gswg2,gswg3} in the context of phononic crystals.
	
	{\it Topological light guiding.---}Inspired by GSWG, which actually is a SPCI with $g=\pm a/2$, we construct L-shaped (U-shaped) SPCIs [see Supplementary Material for the construction details] by shifting PCs along both $x$ and $y$ directions with half lattice constant. We first plot the transmission of GSWG, and L-shaped (U-shaped) SPCIs, as indicated by the black and solid red (blue)  lines, respectively, in Fig.~\ref{Fig_2}(a). It is seen that all three cases exhibit perfect (near unity) transmission within the whole bulk band gap along the SPCIs. Furthermore, the electric field patterns of GSWG, L-shaped, and U-shaped SPCIs in Fig.~\ref{Fig_2}(b-d), respectively, at a frequency of $0.289(c/a)$, show that electromagnetic wave propagate along the L-shape and U-shaped SPCIs with negligible back scattering even in presence of sharp corner, which is more superior than the conventional PC waveguide bending. We emphasize that the perfect transmission of these cases originated from the nontrivial Wannier configuration of the interface, rather than the bulk topology, which differ from the Wannier-type higher-order topological PCs [see the transmission of the Wannier-type higher-order topological PCs in Supplementary Material]. Moreover, we also demonstrate the SPCIs exhibit high fabrication tolerance by implementing numerical simulations in GSWG with disorders [see details in Supplementary Material].
	
\begin{figure}[htbp]
		\centering\includegraphics[width=\columnwidth]{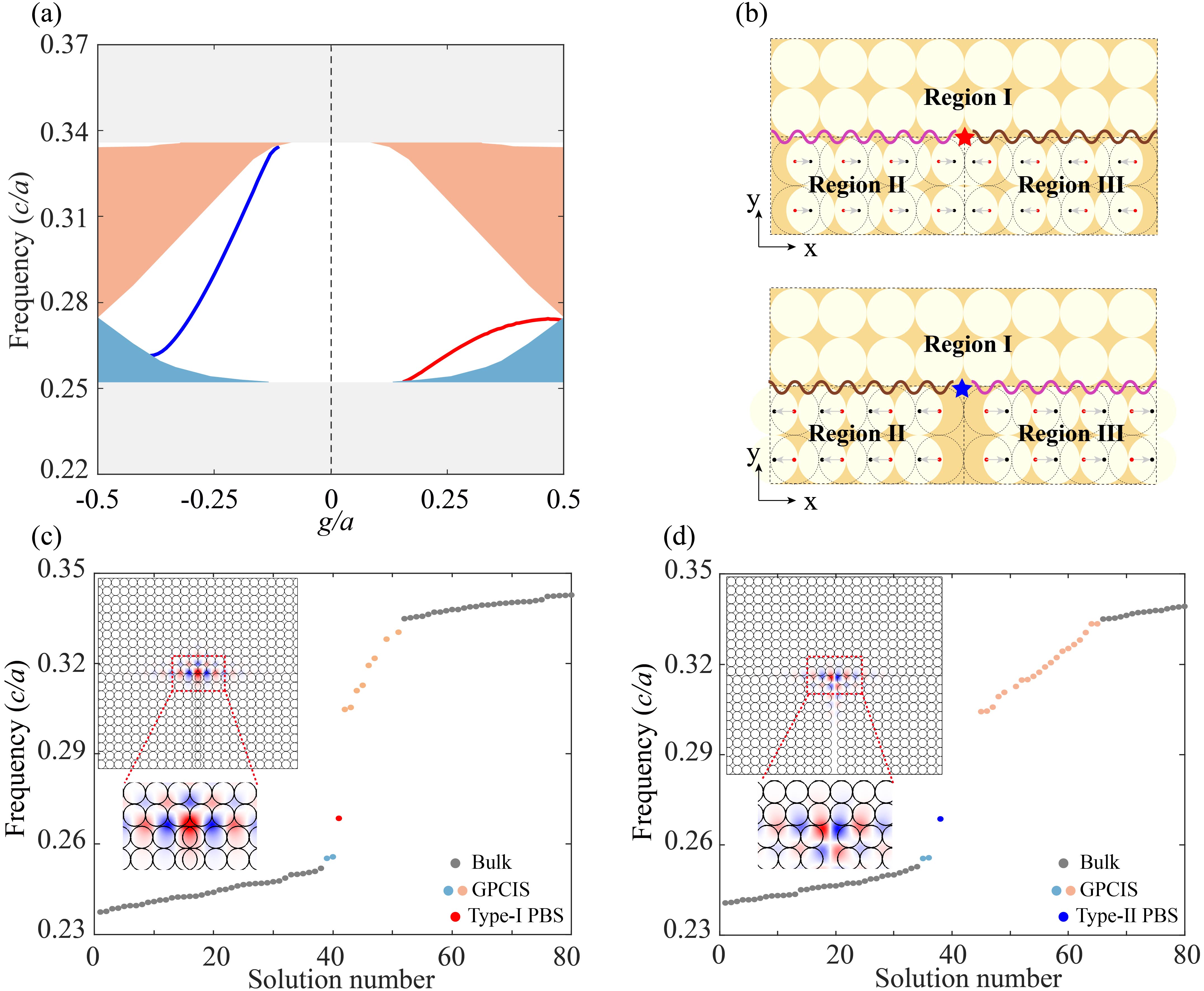}
		\caption{(a) The eigen spectra of the SPCIs versus the shift parameter $g$. The orange and blue areas represent the interface states, while the grey areas refer to the bulk states. The red line refers to the SPCIs-induced PBS. (b) Schematic of the formation of PBS by combining two SPCIs with $|g|$ and $-|g|$ together. Upper panel: the SPCIs with $|g|$ is on the left side of that with $-|g|$, termed as type-I domain-wall system. Lower panel: the SPCIs with $|g|$ is on the right side of that with $-|g|$, termed as type-II domain-wall system. (c,d) Eigen spectra of the (c) type-I and (d) type-II mass domain-wall systems with $g=0.3281a$.}
		\label{Fig_3}
	\end{figure}

	{\it Topological light trapping.---}On the other hand, when $g \neq \pm a/2$, the synthetic Kramers theorem in Eq.~\ref{Kramers} is no longer valid and hence, the degeneracy of two interface bands at $k_x=\pi/a$ will be lifted. As expected, two interface bands separated by a frequency gap emerge in the bulk band gap for the case of $g=0.3a$ [see the dashed colored lines in Fig.~\ref{Fig_1}(d)]. One step further, Fig.~\ref{Fig_3}(b) presents the frequency ranges (indicated by orange and blue areas) of the two projected bands versus the shift parameter $g$. It is observed that the frequency ranges of the two projected interface bands are symmetric about $g=0$ since two SPCIs with opposite $g$ share the identical eigen spectrum. Taking $g \in [0,a/2]$ into consideration, the frequency gap between two projected interface bands is equal to the bulk gap when $g=0$, and gradually decreases and finally closed at $g=a/2$. Meanwhile, the frequency ranges of projected interface bands gradually increase and finally fulfill the whole bulk band gap at $g=a/2$. It is note worthing that the frequency gap between two projected interface bands (described by Dirac mass $m$) experiences a process of the opening and reclosing. Such a Dirac mass $m$ is proportional to the overlapping integral of the electromagnetic field of the two interface states at $k_x=\pi/a$, i.e., $m \propto \int d \bm r( \bm E_+ \cdot {\hat \epsilon}\cdot \bm E_-^\star+c.c.)$, where the subscripts "$\pm$" represent the upper and lower interface bands. Obviously, for $g=\pm a/2$, the parities of $E_\pm$ are well defined, namely, they are of opposite parities about $x=0$ [also see the left panel of Fig.~\ref{Fig_1}(d)], which makes the zero overlapping integral, i.e., $m=0$. Remarkably, a frequency gap with positive (negative) Dirac mass is induced by SPCIs with $g (-g)$ [see details in Supplementary Material]. Hence, one can build a mass domain-wall system by placing two SPCIs with opposite $g (g \neq \pm a/2)$ together. 
	
	In general, there are two configurations of the mass domain-wall systems. As depicted in Fig.~\ref{Fig_3}(b), the first configuration (denoted as type-I, see upper panel) is formed by shifting PCs of region II (III) with positive (negative) $g$, while the second configuration (denoted as type-II, see lower panel) is formed by shifting PCs of region II (III) with negative (positive) $g$. According to the Jackiw-Rebbi theory~\cite{jackiw}, a photonic bound states (PBS) shall be localized at the boundary of two SPCIs with opposite $g$, as schematically indicated by the red and blue stars in Fig.~\ref{Fig_3}(b). Note that Fig.~\ref{Fig_3}(a) also plots the frequency of PBS (indicated by red and blue lines) versus the displacement of PCs between regions I and II. As expected, for both type-I and type-II configurations, the PBS emerges within the frequency gap of the projected interface bands. Specifically, we also present the eigen spectra of type-I and type-II mass domain-wall systems in Figs.~\ref{Fig_3}(c) and (d), respectively, which are formed by SPCIs with $g=\pm 0.3281a$. The corresponding electric field pattern $E_z$ are also shown in the insets. It is seen that the PBS in the type-I mass domain-wall system is even-symmetric, while that with type-II is odd-symmetric. Remarkably, these PBS induced by two SPCIs with opposite $g$ can be acted as photonic cavity modes, which offer an alternative way to realize those cavity-based photonic devices.
	
\begin{figure}[htbp]
		\centering\includegraphics[width=\columnwidth]{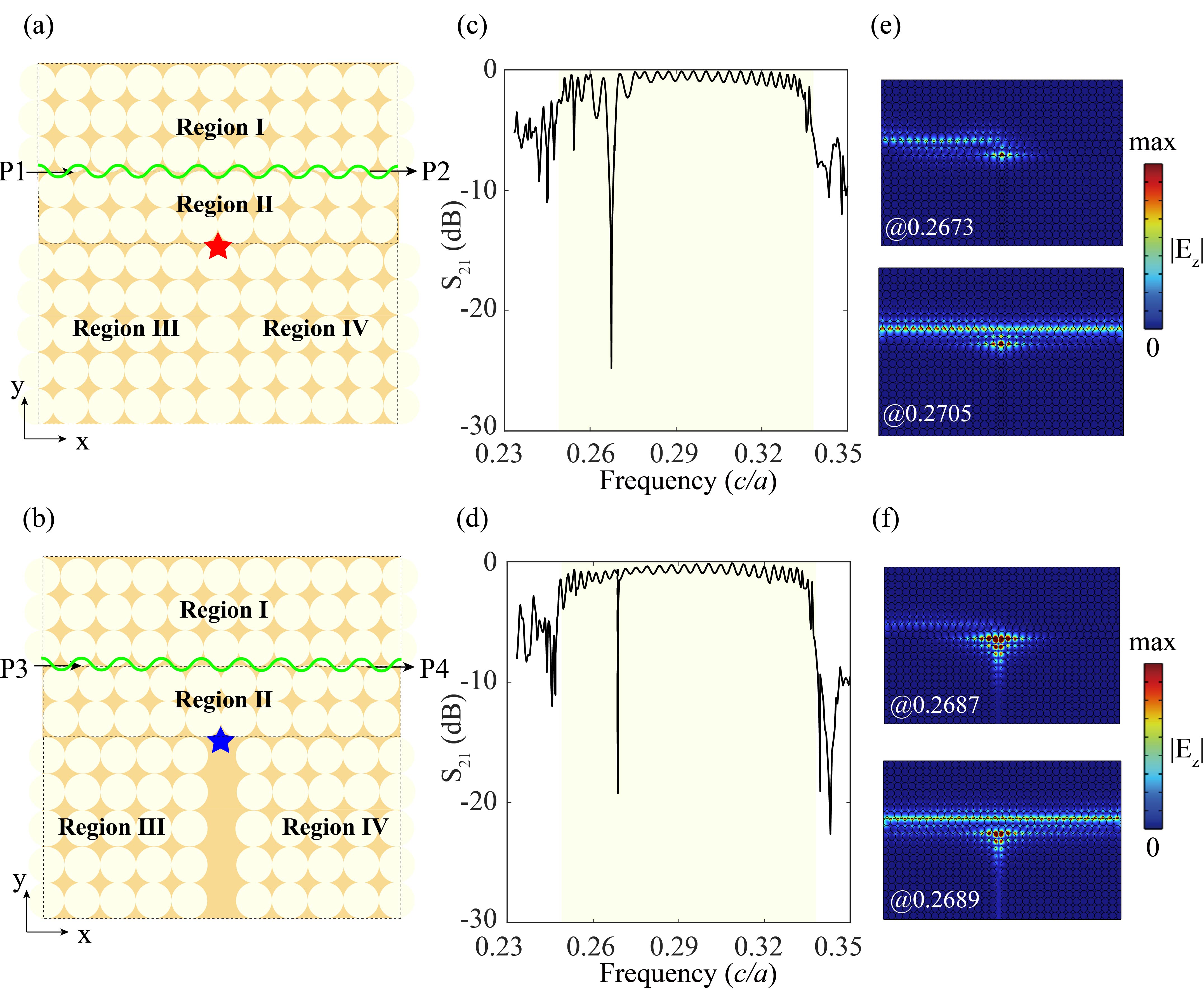}
		\caption{(a,b) Schematic of the formation of (a) type-I, and (b) type-II coupled cavity-waveguide systems. (c,d) Transmission of the (c) type-I, and (d) type-II coupled cavity-waveguide systems. (e,f) The electric field patterns of (e) type-I, and (f) type-II coupled cavity-waveguide systems around the resonance frequencies.}
		\label{Fig_4}
	\end{figure}

	{\it Topological coupled cavity-waveguide systems.---} As an example, we propose a coupled cavity-waveguide (CCW) system by combining the SPCIs-induced waveguide and PBS in a single systems [see Figs.~\ref{Fig_4}(a) and ~\ref{Fig_4}(b)]. Based on the mass domain-wall system with type-I and type-II configurations, we further introduce a GSWG by shifting the PCs of Region I with half lattice constant along $x$-direction and hence, forming type-I (upper panel) and type-II (lower panel) SPCIs-induced CCW systems. The transmission of type-I and type-II CCW systems are displayed in Figs.~\ref{Fig_4}(c) and ~\ref{Fig_4}(d), respectively. It is observed that the transmissions of both systems are nearly unity except for the resonance frequency, which possesses a Lorentzian lineshape. Nevertheless, around the resonance frequency, the half maximum width of transmission of the type-I CCW is much larger than that of the type-II CCW, indicating that Q-factor of the type-I SPCIs-induced PBS is smaller than that of type-II SPCIs-induced PBS. Moreover, we display the normalized electric field patterns of type-I and type-II CCW around the resonance frequency in Fig.~\ref{Fig_4}(e) and ~\ref{Fig_4}(f), respectively.
	
\begin{figure}[htbp]
	\centering\includegraphics[width=\columnwidth]{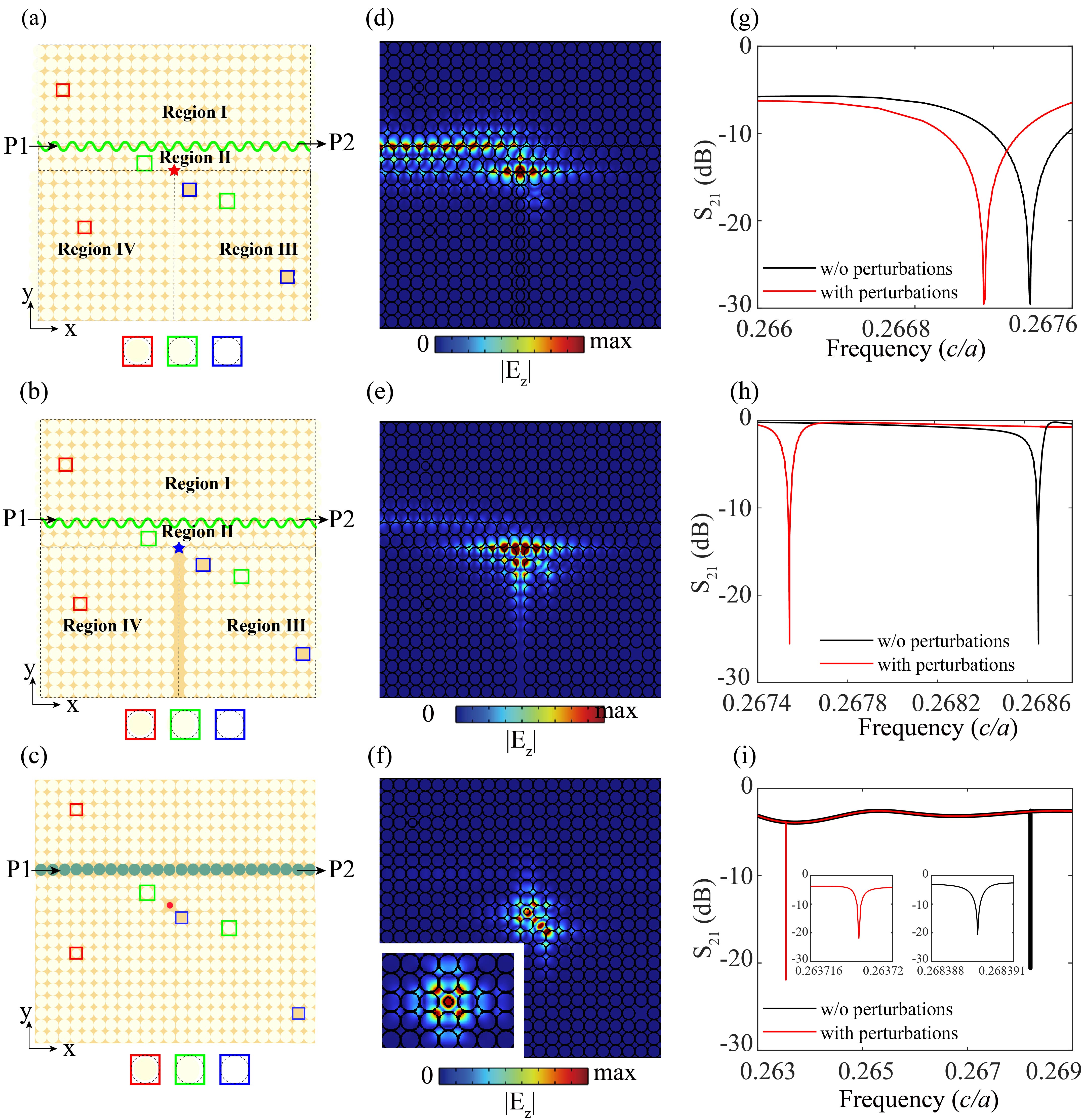}
	\caption{(a-c) Schematic of (a) type-I SPCIs-induced, (b) type-II SPCIs-induced, (c) conventional CCW systems with perturbation. The red, green and blue boxes refer to the radius disorder, location disorder and vacancy defect. (d-f) The normalized electric field patterns at the resonance dip corresponding to (a-c). The inset in (f) refers to the zoom-in normalized electric field pattern at the resonance dip without perturbation. (g-i) The corresponding transmission with (black line) and without (red line) perturbations of (a-c) around resonance frequency. Inset of (i) refers to the zoom-in transmission lineshape around the resonance frequencies.}
	\label{Fig_5}
\end{figure}

Last but not least, it is note worthing that the CCW systems formed by SPCIs are robust against perturbations. As shown in Figs.~\ref{Fig_5}(a) and ~\ref{Fig_5}(b), typical perturbations, including two radii disorders, two locations disorders, and a vacancy defect, which indicated by the red, green and blue boxes, respectively, are randomly introduced into the type-I and type-II SPCIs-induced CCW systems. As a comparison, we also introduce the same perturbations into a conventional CCW system based on the square PC [see Fig.~\ref{Fig_5}(c)], in which the cavity and waveguide are formed by revising the radius of an air hole to $0.229a$ (see the red circle) and replacing one row of the air holes with a row of the dielectric rods (the permittivity of the replaced rods is $4.2$, see the dark green circles). It is seen that for both type-I and type-II SPCIs-induced CCW systems with perturbations, the normalized electric field patterns in Figs.~\ref{Fig_5}(d) and ~\ref{Fig_5}(e) are almost remain unchanged compared to the cases that without perturbations in Figs.~\ref{Fig_4}(e) and ~\ref{Fig_4}(f), implying that the coupling at resonant dip does not interrupted by nearby defects. In contrast, for the conventional CCW system with perturbations, the normalized electric field patterns in Fig.~\ref{Fig_5}(f) at the resonance dip makes a huge difference with the case that without perturbations [see the inset of Fig.~\ref{Fig_5}(f)], indicating that the cavity mode is destroyed by the defect.
 
Moreover, we display the transmission of the type-I and type-II SPCIs-induced and conventional CCW systems with and without perturbations around the resonance dip in Fig.~\ref{Fig_5}(g-i). For all these cases, resonance dip experience a blue shift due to the introducing of the defects. Nevertheless, the frequency shifts of the type-I and type-II SPCIs-induced CCW, namely $0.0002(c/a)$ and $0.0011(c/a)$, are much smaller than that of the conventional CCW, namely $0.0047(c/a)$, indicating the resonance frequency of the SPCIs-induced CCW is also stable against perturbations. The robustness of the SPCIs-induced CCW against both disorder and defect make it a potential candidate for high-performance building blocks of miniature integrated topological photonic circuits.

	{\it Conclusion.---} To conclude, the SPCIs that simply formed by gliding the PC are proposed to guide and trap light. The SPCIs are closely rely on the shift parameter $g$. We demonstrate that SPCIs with $g=\pm a/2$ forms a GSWG and exhibit excellent performance on the ability of guiding waves even in the presence of sharp bends and disorders. Moreover, placing two SPCIs with opposite Dirac mass results in a PBS due to the Jackiw-Rabbi mechanism, which can be acted as a photonic cavity mode. Our work provides an alternative way towards the design of ultracompact integrated photonics. 
	\\
	
	{\it Acknowledgement.---}This work is supported from the Natural Science Foundation of Guangxi Province (Grant No. 2023GXNSFAA026048),and the project of all-English course construction for graduate students in Guangxi Normal University.

\end{document}